\documentstyle[12pt,leng]{article}
\begin{document}
\hfill{LA-UR-93-2406}

\vspace{7pt}
\begin{center}
{\large\sc{\bf Bosonization
of $QED_3$ with an induced Chern - Simons term.}}

\baselineskip=12pt
\vspace{35pt}

A. Kovner$^*$ and P. S. Kurzepa $^{**}$\\
\vspace{10pt}
Theory Division, T-8,
Los Alamos National Laboratory,
MS B-285\\
Los Alamos, NM 87545\\
\vspace{55pt}
\end{center}
\begin{abstract}
We extend the bosonization of $2+1$ -
dimensional QED with one fermionic flavor performed previously to the
case of QED with an induced Chern - Simons term.
The coefficient of this term is quantized:
$e^2n/8\pi$, $n\in {\bf Z}$. The fermion operators are
constructed in terms of the bosonic fields
$A_i$ and $E_i$. The construction is similar to that in the $n=0$ case.
The resulting bosonic theory is Lorentz invariant in the continuum limit and
has Maxwell's equations as its equations of motion.
The algebra of bilinears exhibits nontrivial operatorial mixing
with lower dimensional operators, which is absent for $n=0$.

\end{abstract}

\vspace{50pt}
\
\newline
*KOVNER@PION.LANL.GOV
\newline
**KURZEPA@MUON.LANL.GOV
\vfill
\pagebreak

Recently, the complete bosonization of 1 - flavor $QED_3$ has been
performed \cite{boson} through a direct
extension of Mandelstam's procedure \cite{mandelstam}
to higher dimensions. This is the first instance, in
which a {\it relativistically invariant} $2 + 1$ - dimensional theory
has been bosonized operatorially in terms of {\it local} bosonic variables.
It is not clear yet, whether the bosonization procedure described
in \cite{boson} is as general as its 1+1 dimensional analog
and can be applied to any 2+1 - dimensional fermionic model.
It would be therefore interesting to extend the results of \cite{boson}
to other $2 + 1$ - dimensional theories.

In this note we make a modest
step in this direction. We consider QED$_3$ with an induced Chern - Simons
term \cite{deser}. It is well known \cite{coste} that QED$_3$ has a peculiar,
anomaly - like feature. The low energy behaviour of the theory
is determined by a Chern - Simons term, whose coefficient depends
on the way one regularizes the theory in the ultraviolet.
In \cite{boson} we considered only the regularization which
does not lead to the appearance of an
induced Chern - Simons term. In the present paper
we address the question of regularization dependence, and perform the
bosonization of QED$_3$ with
an arbitrary value of the Chern - Simons coefficient which can be
induced by inequivalent ultraviolet regularizations of the theory.
The resulting bosonic theory is local, and Lorentz invariant,
and the results of \cite{boson} are recovered in the appropriate limit.
It is interesting to note that our bosonization procedure can be directly
applied {\it only} for those values of the coefficient of the Chern - Simons
term which can be interpreted as appearing entirely due to the regularization
ambiguity mentioned above. For other values of the coefficient, the
bosonic theory which we obtain is nonlocal. It is not clear to us
whether the procedure can be modified to incorporate also other values.

The appearance of the Chern - Simons term in $QED_3$ is best
understood by first considering Dirac fermions in {\it external}
electromagnetic fields.
As is well known \cite{gordon}, the determinant of the $2 + 1$ - dimensional
Dirac operator with an external Abelian gauge field cannot be defined uniquely.
This discrete ambiguity is, in fact, determined by the assumed ultraviolet
behavior of the Dirac propagator, which is used to compute this functional
determinant.
That the ambiguity is discrete is most easily seen in the framework of the
lattice - regularized theory with Wilson fermions \cite{coste}.
The behavior of the Wilson propagator
near the top edge of the Brillouin zone, ${\bf T^3}$, may be modified without
changing the continuum limit only if the winding number of this propagator,
regarded as a map fron ${\bf T^3}$
into nonsingular 2 - dimensional matrices, is integer.

The ambiguity results in the appearance of a Chern - Simons
term with a {\it quantized}
coefficient in the effective action for the Dirac fermion
in the external field. Now, when considering $QED_3$, i. e., quantizing also
the gauge field, one has two options. The first is to modify the ultraviolet
behavior of the fermion propagator consistently with the desired value of
the Chern - Simons coefficient. However, since the {\it only} effect
of this modification is to induce a local Chern - Simons term,
one can alternatively include this term explicitly in the
{\it classical} action. In this paper we will use the second option, which
enables us to use the same ultraviolet regularization as in \cite{boson}.

Accordingly, we consider the following Lagrangian,
\begin{equation}
L  = -\frac{1}{4} F^{\mu \nu}F_{\mu \nu} -  \lambda\epsilon ^{\mu \nu \alpha }
 F_{\mu \nu } A_{\alpha} + \bar \chi \gamma^i (\partial_i + ieA) \chi - m \bar
\chi \chi
\end{equation}
where,
\begin{equation}
\lambda = \frac{e^2}{8 \pi} n; \ \ n\in {\bf Z}
\label{lambda}
\end{equation}
and $\chi$ is a 2 - component Dirac spinor.
The corresponding Hamiltonian [we will work in the temporal
(Weyl) gauge $A_0 = 0$] is,
\begin{equation}
H=\frac{1}{2}E^2+\frac{1}{2}B^2+\bar\chi\gamma^i(i\partial_i+eA_i)\chi+m\bar\chi \chi
\label{ham}
\end{equation}
The symplectic structure of the theory is modified due to the
presence of the
Chern - Simons term.  The momentum conjugate to $A_i$
is now $p_i$,
\begin{equation}
p_i = -E_i + \frac{e^2 n}{4 \pi}\tilde A_i
\end{equation}
with $\tilde A_i \equiv \epsilon_{ij}A_j$.
The canonical commutators are,
\begin{equation}
[ E_i(x) ; A_j(y) ] = i \delta_{ij}\delta (x - y),\ \
[E_i(x) ; E_j(y)] =  \break -in \frac{e^2}{2 \pi} \epsilon_{ij}\delta (x - y)
\end{equation}
It is convenient for future use to introduce the field $\Pi_i(x)$,
\begin{equation}
\Pi_i = E_i - \frac{e^2 n}{2 \pi}\tilde A_i
\end{equation}
which satisfies $ [\Pi_i(x); \Pi_j(y)] = in \frac{e^2}{2 \pi}\epsilon_{ij}
\delta (x - y) $.
Gauss' constraint, to be appended to the
Hamiltonian eq.( \ref{ham}), is,
\begin{equation}
\partial_i \Pi_i = e \chi^\dagger\chi
\label{const}
\end{equation}

Our aim is to construct a Dirac doublet of fermionic operators
in 2+1 dimensions in terms of {\it local} bosonic fields.
To construct the operator $\chi$ we must first fix the gauge freedom
associated with the time independent gauge transformations, generated
by the Gauss' constraint. We do this by considering $\chi$ in the
Coulomb gauge. When written in terms of these operators, the
covariant derivative in eq.(\ref{ham})
contains only the transverse part of the vector potential $A_i$.
We shall work with the gauge - invariant operators,
\begin{equation}
\chi^{CG}_\alpha(x)
=\chi_\alpha(x)\exp{\left[ie\int d^2y e_i(y - x)A_i(y)\right]}
\label{cg}
\end{equation}
where,
\begin{equation}
e_i(x)=-\frac{1}{2\pi}\frac{x_i}{x^2}
\end{equation}
is the electric field of a point
charge. (We will drop the superscript $CG$ from now on).

We will now solve Gauss' constraint by constructing
the doublet of anticommuting fermionic operators $\chi_\alpha$
in terms of the local fields $E_i$ and
$A_i$.
Substituting those into the Hamiltonian,
eq.(\ref{ham}), we will obtain the bosonized form of
Chern - Simons $QED_3$,
defined on the physical Hilbert space.
We will then calculate the fermionic bilinears, and
establish Lorentz invariance.

Our construction initially
follows the same steps as  in the $n = 0$ case. We first
define the anticommuting operators $\psi_\alpha(x)$ in terms of the charged
field,
\begin{equation}
\Phi(x)=exp\left[ie\int d^2y e_i(y - x)A_i(y)\right]
\end{equation}
and the operators $V_\alpha(x)$ ($U_\alpha(x)$) which create (annihilate)
a magnetic vortex of a half -  integer strength,
\begin{equation}
V_1(x)=
-i\exp\left\{\frac{i}{2e}\int d^2y\left[(\theta(x-y)-\pi)\partial_iE_i(y)
+2\pi G^{(2)}(y-x)\epsilon_{ij}\partial_iE_j(y)\right]\right\}
\label{vu}
\end{equation}
$$U_1(x)=
\exp\left\{-\frac{i}{2e}\int d^2y\left[\theta(y-x)\partial_iE_i(y)+
2\pi G^{(2)}(y-x)\epsilon_{ij}\partial_iE_j(y)\right]\right\}$$
$$V_2(x)=-iV^\dagger_1(x); \ \ \ \ U_2(x)=U^\dagger_1(x)$$
Here $G^{(2)}(x-y)=-\frac{1}{4\pi}\ln( x^2\mu^2)$, and $\theta$ is an angle
function, constructed in \cite{boson}. The interpretation
of $V$ and $U$ as vortex operators follows from the following
commutation relations,
\begin{equation}
[V_1(x),B(y)]=-\frac{\pi}{e}V_1(x)\delta^2(x-y) ;\ \
[U_1(x),B(y)]=\frac{\pi}{e}U_1(x)\delta^2(x-y)
\end{equation}

The regularized  fermionic operators $\psi_\alpha$ are defined as follows,
\begin{equation}
\psi^\eta_1(x)=
V_1(x+\eta)\Phi(x)U_1(x-\eta); \ \ \psi^\eta_2(x)= V_2(x+\eta)\Phi(x)
U_2(x-\eta)
\label{psieta}
\end{equation}

The anticommutativity of these operators when the separation between
theeir arguments is much larger than the length of the regulator
($|x-y|>>|\eta|$) can be checked by a direct calculation \cite{boson}.
The anticommutativity of $\psi_1(x)$ with $\psi_1(y)$ and $\psi_2(x)$
with $\psi_2(y)$ follows from the following property of
the angular function $\theta$, $\theta(z)-\theta(-z)=\pm\pi$. The
anticommutativity of $\psi_1(x)$ with $\psi_2(y)$ is ensured by
the relative factor,
\begin{equation}
N=exp[-\frac{i\pi}{e}\int d^2y\partial_iE_i(y)]
\label{klein}
\end{equation}
in the definitions of $V_1$ and $V_2$ in eq.(\ref{vu}).
The factor $N$ thus plays the same role as the Klein factor \cite{halpern} in
1+1
dimensions.

At this point we must depart slightly from the $n=0$ case.
The fermionic operator $\chi_\alpha(x)$, as defined in eq.(\ref{cg}),
has a local commutator with the topological charge density
$[\partial_iE_i(x),\chi_\alpha(y)]=-e\chi_\alpha(y)\delta^2(x-y)$.
Since, however,
\begin{equation}
\left[ \partial_i E_i (x) ; \int d^2z G^2(y-z) \epsilon_{ij}
\partial_i E_j (y - z)\right] = i \frac{n e^2}{2 \pi} \delta^2(x - y)
\end{equation}
this is not true for $\psi_\alpha^\eta(x)$.
This can be remedied by
modifying the expressions in eq.(\ref{psieta}) by additional factors
in such a way that the products of the
vortex operators $V( x + \eta ) U( x - \eta)$, together with these factors,
commute with $\partial_i E_i$.

In fact, there is also an additional reason
for this modification. It is
intuitively clear that
the basic ingredients of our contruction should be closely
related to the local dual variables of QED$_3$ \cite{marino}. These, however,
carry magnetic flux, but no topological charge, measured by $\partial_i E_i$.
This is, in fact, a necessary condition
for their locality: any operator that carries a nonzero
topological charge is necessarily nonlocal relative to electric field,
which itself is a local gauge invariant
operator\footnote{Although the operators $V_\alpha$ and $U_\alpha$
carry half integer flux, and thus are not local and not single valued
themselves, they can be thought of as "square roots" of a
single valued operator ${\cal V}$. This field ${\cal V}$
carries an integer magnetic flux and is the local dual field,
since $V_\alpha$ and $U_\alpha$ as modified below do not carry topological
charge.}.

We therefore modify the fermionic operators by multiplying them by
the following Wilson factors,
\begin{equation}
W_{1, 2}[ L ; x, \eta ]
= exp{ \left[ \mp ie \frac{n}{2}\int_L dl_i A^i \right]}
\end{equation}
Here $L$ is the straight line connecting $x - \eta$ and $x + \eta$.
The modified operators are,
\begin{equation}
\chi_{1,2}^\eta(x)=\psi_{1,2}^\eta(x)W_{1, 2}[ L ; x, \eta ]
\end{equation}
The Wilson lines have the following desirable feature. Since,
\begin{equation}
\left[\int d^2 z G^{(2)}(y - z)
\epsilon_{ij} \partial_i E_j ;
\int_L dl_i A^i\right] = - i\int_L dl_i \partial^i \Theta ( y - z )
\label{adphase}
\end{equation}
this commutator vanishes in the limit $| x - y | >> \eta$.
Therefore the anticommutation relations of the $\chi_\alpha$'s at distances
much larger than the regulator are the same as those of the $\psi_\alpha$'s.

In the final expression for the fermionic operators we must
average over the direction of the regulator $\eta$, which implements the
point splitting. The averaging is performed with appropriate phase factors
to ensure the correct rotational properties of the $\chi_\alpha$'s,
\begin{equation}
\chi_1(x)=\lim_{\Lambda\rightarrow \infty}\frac{k\Lambda}{2\pi }
\int d\hat\eta e^{-i\frac{\theta(\eta)}{2}}\chi_1^{\eta}(x) ; \ \
\chi_2(x)=\lim_{\Lambda\rightarrow \infty}\frac{k\Lambda}{2\pi }
\int d\hat\eta e^{i\frac{\theta(\eta)}{2}}\chi_2^{\eta}(x)
\label{psi}
\end{equation}
Here k is a number,
depending on the precise definition of the ultraviolet cutoff $\Lambda$,
the length of the regulator $\eta$ is taken to be proportional to the
inverse of the ultraviolet cutoff, $|\eta|\propto 1/\Lambda$, and
the integral is
over the angle of the vector $\eta$, $\hat\eta$.

We can now easily check, that
the fermionic operators $\chi_\alpha$ satisfy the following
conditions,

i. $[\chi_\alpha(x),\partial_iE_i(y)]=e\chi_\alpha(x)\delta^2(x-y)$, i. e.,
carry unit electric charge,

ii. Under a rotation by $\phi$ \footnote{We are
using the following basis of the Dirac matrices:
$\gamma^0=\sigma^3 ;\  \gamma^1=i\sigma^2 ;\ \ \gamma^2=-i\sigma^1$.},

\begin{equation}
\chi_1\rightarrow e^{i\phi/2}\chi_1; \ \ \chi_2\rightarrow e^{-i\phi/2}\chi_2
\label{rot}
\end{equation}

iii. In the limit $ \Lambda \rightarrow \infty $ have
the correct anticommutation relations.

We can now calculate the fermionic bilinears. Those should also
be defined using the point splitting procedure,

\begin{equation}
J_\Gamma(x)=\bar\chi(x)\Gamma\chi(x)\equiv \frac{1}{8\pi}
\int d\hat\epsilon e^{i\omega_\Gamma(\hat\epsilon)}\left\{\left[
\chi^\dagger(x+\epsilon),\gamma^0\Gamma\chi(x-\epsilon)\right]
,e^{ie\int_{x-\epsilon}^{x+\epsilon}dx_i
A_i^{tr}}\right\}_{|\epsilon|\propto|\eta|}
\label{bilinear}
\end{equation}

It is understood in eq.(\ref{bilinear}) that the
limit $\Lambda\rightarrow \infty$ is taken after
(independent) averaging over the directions of $\epsilon$ and $\eta$.
Since we are constructing the
Coulomb gauge fermions, it is only the transverse part of the vector
potential, $A^{tr}$, which appears in the additional Wilson factor. The phase
$\exp\{i\omega_\Gamma\}$ is inserted to project onto
the relevant irreducible representation of the 2D
rotation group while averaging over $\hat \epsilon$.
Thus, for $\Gamma=\gamma_0$ and $\Gamma=1$ we have
$\omega(\epsilon)=0$, while for $\Gamma=\gamma_+\equiv\gamma_1+i\gamma_2$,
$\omega(\epsilon)=\theta(\hat\epsilon)$, and for $\Gamma=\gamma_
-\equiv\gamma_1-i\gamma_2$, $\omega(\epsilon)=-\theta(\hat\epsilon)$.
We normal - order the bilinears with respect to the
perturbative vacuum.

We have calculated the bilinears using the definitions
of the fermionic operators given above.  The calculation is performed,
in analogy to the 1+1 dimensional case, by expanding all quantities
in powers of the inverse cutoff. Since the computational strategy is
essentially the same as in the $ n = 0 $ case, we refer the reader
to the paper \cite{boson} for the missing details. Here we wish only to stress
three salient points.

It is too naive to neglect in the expansion higher - order operatorial terms
multiplied by inverse powers of $\Lambda$. These are formally small, but
give, in fact, finite contributions, if these operators have high
enough dimensions. This problem
arises also in 1+1 dimensions, where the effect of the higher - order terms is
to renormalize the coefficients of the lower dimensional operators in a way
consistent with the symmetries of the problem \cite{mandelstam}.

In order to perform the actual calculations, we must understand the
ultraviolet behavior of the bosonized theory.
This behavior is different from the
naive one, obtained in perturbation theory. This is related to the fact
that, apart from the
ultraviolet cutoff $\Lambda$, the bosonized
theory has an additional ultraviolet scale $\mu = (e^2\Lambda^2)^{1/3}$.
This scale appears since the fermionic operators we have constructed
are not pointlike. At distances smaller than $1/\mu$
the non - point - likeness of $\chi$ becomes important.
Dimensional considerations, utilizing the
bosonized Hamiltonian, show \cite{boson} that above the scale
$\mu$ there is a crossover in
the scaling behaviour of the electric field, $<E_i(x)E_i(y)>_{|x-y|>>1/\mu}
\sim \frac{1}{|x-y|^3}$, while
$<E_i(x)E_i(y)>_{|x-y|<<1/\mu}
\sim \frac{e^2}{|x-y|^2}$.
At finite $e^2$ this additional scale
becomes infinite in the continuum limit
$\Lambda\rightarrow\infty$, and therefore
the usual scaling behaviour of perturbative $QED$ is recovered at all scales.
However, in the process of calculation of the bilinears we
have to perform operator product expansions near the scale of the
UV cutoff $\Lambda$. When working in this regime,
we take the following form of the utraviolet asymptotics
of the correlators of the bosonic fields, which is the most general one
consistent with this ultraviolet scaling dimension, rotational symmetry,
and parity transformation properties,
\begin{equation}
lim_{x\rightarrow y}<A_i(x)A_j(y)>=r_1\frac{\delta_{ij}}
{|x-y|}+2r_2\frac{(x-y)_i(x-y)_j}{|x-y|^3};
\label{correl}
\end{equation}
$$
lim_{x\rightarrow y}\frac{1}{e^2}<E_i(x)E_j(y)>=q_1
\frac{\delta_{ij}}
{|x-y|^2}+2q_2\frac{(x-y)_i(x-y)_j}{|x-y|^4};
$$
$$
lim_{x\rightarrow y}<E_i(x)A_j(y)>=s_1\frac
{\delta_{ij}}{|x-y|^2}+2s_2\frac{(x-y)_i(x-y)_j}{|x-y|^4}
$$
$$
+\hat mp_1\frac{\epsilon_{ij}}{|x-y|}+2\hat mp_2\frac
{(\tilde x-\tilde y)_i(x-y)_j-
(\tilde x-\tilde y)_j(x-y)_i}{|x-y|^3};
$$
with  $\hat m=m+e^2nu$ where $u$ is a numerical coefficient.
This form is the same (with the possible exception of the values of
the numerical coefficients multiplying the various terms), as in the
$n = 0$ case.

When calculating the coefficients of the $1/\Lambda$ expansion, various
averges over the regulator directions, discussed extensively
in \cite{boson}, have to be performed. These averages involve phase factors,
inherent in the definition of the fermionic bilinears, and the fermionic
operators themselves. We shall not discuss this issue further here, but
note only that all the phase factors remain unchanged from the $n = 0$
case. This is because the additional phase in eq.(\ref{adphase}) tends
smoothly to $0$ for $|\eta| \rightarrow 0$. Thus, for $|\epsilon| >> |\eta|$,
which is the appropriate regime for averaging,
it does not contribute to the averaging. In fact, it can be immediatelly seen
that the only modification from the $n=0$ case is that in all expressions
for the fermionic bilinears, the electric field $E_i$
has to be replaced by $\Pi_i$.

We now quote the results of the calculations of the bilinears in
Chern - Simons $QED_3$,

\begin{equation}
J_0\equiv\chi^\dagger\chi=\frac{1}{e}\partial_i \Pi_i
\label{charge}
\end{equation}

\begin{equation}
J\equiv:\bar\chi\chi:=-2:A_i\tilde\Pi_i:
\label{mass}
\end{equation}

The calculation of the spatial components of the current is more
involved. Here one encounters the problem mentioned earlier,
namely, that higher order terms in the naive expansion in powers of the inverse
cutoff cannot be neglected. It can be shown, however, that their effect
is only to renormalize the coefficients of the lower order terms \cite{boson}.
The exact form of the current
can be determined by requiring that they satisfy the
tree - level current algebra
(for $n=0$), possibly modified by Schwinger terms, and terms
that vanish in the continuum limit, and by the requirement of the restoration
of Lorentz invariance in the continuum limit. The situation
here is analogous to that in 1+1 dimensions
\cite{mandelstam}, where the use of Lorentz invariance is required
to fix the overall scale of the current.
The current is then found as\footnote{Here
and throughout the paper the symmetric, or Weyl,
ordering of $A_i$ and $\Pi_i$ is understood in all of the expressions
for composite operators.} ,
\begin{equation}
J_i\equiv\bar\psi\gamma_i\psi= - e\kappa\Lambda A_i+
\frac{1}{e\kappa\Lambda}\left[2\tilde \Pi_i(A \tilde \Pi)
- \Pi^2 A_i \right] + \frac{M}{e}\tilde \Pi_i
\label{Current}
\end{equation}
The constants $\kappa$ and $M$  are expressed in terms
of the numerical coefficients which appear
in eq.(\ref{correl}) or, alternatively, in terms of
the vacuum expectation values of the bosonic operators,
\begin{equation}
\kappa=\frac{\Lambda}{e^2}\frac{<\Pi^2>}{<B^2>}; \ \
M=-\frac{2}{\kappa\Lambda}<A\cdot\tilde \Pi>
\label{consta}
\end{equation}
The leading term on the right hand side of eq.(\ref{Current})
is proportional to the ultraviolet cutoff. We
therefore keep also the next order term, which vanishes in
the naive continuum limit.
The fermionic bilinears are local
functions of $\Pi_i$ and $A_i$.

The algebra of the bilinears is,
\begin{equation}
[J_i(x) , J_j(y)]=2i\epsilon_{ij}J(x)\delta^2 (x-y)
\label{algebra}
\end{equation}
$$
 -i\frac{n}{2\pi}\left\{
\frac{4}{\kappa^2\Lambda^2}[(A\Pi)^2+(A\tilde\Pi)^2]+\frac{4M}{\kappa\Lambda}
A\tilde\Pi+M^2\right\}\epsilon_{ij}\delta^2(x-y);
$$
$$
[J_i(x),J(y)]=-2i\epsilon_{ij}J_j(x)\delta^2(x-y)
+in\frac{e^2}{\pi}\left [ \frac{M}{e}\tilde A_i +  \frac{1} {e\kappa \Lambda}
\Pi_i A^2\right]\delta^2 (x - y);
$$
$$
[J_0(x) , J_i(y)]=-i\left[\Lambda \kappa \delta_{ij}+
\frac{2}{e^2\Lambda\kappa}(\Pi_i\Pi_j-1/2\Pi^2\delta_{ij})
\right]\partial^x_j\delta^2 (x-y)
$$
$$
+ i\frac{n}{2\pi\kappa\Lambda}\left [ (2A\tilde\Pi +\kappa M\Lambda)
\delta_{ij}+2A\Pi\epsilon_{ij} \right ] \partial_j^x \delta^2 (x - y);
$$
$$
[J(x) , J_0(y)] = -\left[ \frac {2i}{e}
\tilde \Pi_i (x) + n \frac {ie}{\pi} A_i(x)\right] \partial_i^{x}\delta^2 (x -
y)
$$

Note that while the fermion currents $J_i$ and $J$
in the free theory (as well as for $n = 0$)
obey the $SU(2)$ algebra, this is no longer true here.
Thus, for instance, performing the operator product expansion with
the asymptotics of eq.(\ref{correl})
for the operator in the curly brackets in the first line of eq.(\ref{algebra}),
we find,
\begin{equation}
\frac{4}{\kappa^2\Lambda^2}[(A\Pi)^2+(A\tilde\Pi)^2]+\frac{4M}{\kappa\Lambda}
A\tilde\Pi+M^2=2e^2\gamma A^2+O(1/\Lambda)
\end{equation}
with,
\begin{equation}
\gamma = - \frac{2<B^2>}{\Lambda^3 \kappa}
\label{gamma}
\end{equation}
The commutator of
the spatial components of the current involves therefore, in addition to the
mass term,  the lower - dimensional operatorial term $A^2$.
This is a novel example
of operator mixing, allowed in the asymptotically free Chern - Simons
$QED_3$. In the $n = 0$ theory such additional terms are excluded by parity
considerations. The same is true for all the terms in eq.(\ref{algebra}) which
vanish at $n=0$. They all represent operatorial mixing with
the lower - dimensional operators which are not allowed by parity at $n=0$, and
vanish smoothly in the limit of the free theory $e^2\rightarrow 0$.
The commutators are also modified from their tree - level forms by
Schwinger terms. These terms are, in fact, necessary to maintain the
Lorentz invariance of the quantized theory.

The analogous calculation of the bosonized form of the energy - momentum
tensor gives,

\begin{equation}
T^{00}=\frac{1}{2}B^2+\frac{1}{2}E^2+
\frac{1}{2e^2\kappa\Lambda}(\partial_i\Pi_i)^2+\frac{e^2}{2}\kappa\Lambda A^2
\label{fin}
\end{equation}
$$
+\frac{1}{2\kappa\Lambda}\left[-2(A\cdot\tilde \Pi)^2+
\Pi^2A^2 \right]+M A\cdot\tilde \Pi;
$$
\begin{equation}
T^{0i}= B\tilde E_i - (\partial_j \Pi_j) A_i +\gamma
\partial_i(A\Pi) + (\frac{M}{8}+
\frac{e^2}{8\pi}n\gamma )\partial_j(A_i \tilde A_j + A_j \tilde A_i) +
O(\frac{1}{\Lambda})
\label{momentum}
\end{equation}
with $\kappa$, $M$ and $\gamma$ given in eqs. (\ref{consta}), (\ref{gamma}).

Calculating the commutators $[ B , H ]$, and $[ E , H ]$, we recover
Maxwell's equations of the Chern - Simons theory up to terms which vanish in
the continuum limit.
Finally, Lorentz invariance of the theory is verified by checking that
Schwinger's commutator condition \cite{schwinger},
\begin{equation}
-i[T_{00}(x),T_{00}(y)]=-(T^0_i(x)+T^0_i(y))\partial^i\delta^2(x-y)
\label{schwinger}
\end{equation}
holds up to terms of order $1/\Lambda$.

We have determined all the coefficients in
the Hamiltonian density and the currents
in terms of the expectation values $<\Pi^2>$, $<B^2>$, and $<A\cdot\tilde
\Pi>$.
The coefficients $\kappa$, $M$,
and $\gamma$, which enter the expressions for
the energy - momentum tensor, are therefore {\it not} free paprameters. They
are adjusted dynamically to satisfy eqs.(\ref{consta}),
(\ref{gamma}). However, since the bosonic theory defined by the
Hamiltonian eq.(\ref{fin}) is
strongly interacting,
we cannot determine these expectation values analytically.

We have expressed the one flavor $QED_3$ with an induced  Chern - Simons term
in terms of the  canonical bosonic fields $A_i$ and $E_i$.
The resulting theory is local, and Lorentz invariant. The
fermionic operators and their bilinears obtained become,
in the limit $n \rightarrow 0$,
identical to those previously calculated in $QED_3$.

Finally we note that our present method leads to a local bosonic
theory only for the quantized values of the Chern - Simons coefficient
given in eq.(\ref{lambda}). The reason is the following.
Throughout the calculation one encounters the square of the Klein factor
N of eq.(\ref{klein}).
For the values of $\lambda$ of eq. (\ref{lambda})
both the fermion number operator, $\frac{1}{e}\int d^2y\partial_i\Pi_i(y)$,
and the topological charge operator,
$\frac{1}{e}\int d^2y\partial_iE_i(y)$, have integer eigenvalues.
Therefore, as in 1+1 dimensions,
the Klein factor $N$ has eigenvalues $\pm 1$ and $N^2=1$. If this
quantization condition does not hold, $N^2\ne 1$ and $N^2$
is a nonlocal operator.
The spatial components of the current and the energy density are
then multiplied by a nonlocal operator of this type.
We are currently investigating the question, whether the construction
can be modified to incorporate arbitrary Chern - Simons coefficients.

\end{document}